\documentclass[epsfig,12pt]{article}
\usepackage{epsfig,amssymb,euscript}
\usepackage{amsmath}

\newcommand{\vol}{\mathrm{vol}}
\newcommand{\mz}{\mathbb{Z}}

\newcommand{\be}{\begin{equation}}
\newcommand{\ee}{\end{equation}}
\newcommand{\bea}{\begin{eqnarray}}
\newcommand{\eea}{\end{eqnarray}}
\newcommand{\ba}{\begin{array}}
\newcommand{\p}[1]{(\ref{#1})}
\newcommand{\ea}{\end{array}}

\def\bbox{{\,\lower0.9pt\vbox{\hrule \hbox{\vrule height 0.2 cm
\hskip 0.2 cm \vrule height 0.2 cm}\hrule}\,}}
\newcommand{\dsl}{\pa\kern-0.5em /}

\newcommand{\nn}{\nonumber \\}

\newcommand{\de}{\partial}

\def\ds{\raise.15ex\hbox{/}\kern-.57em\partial}
\def\Ds{\,\raise.15ex\hbox{/}\mkern-13.5mu D}

\newcommand{\bZ}{\mathbb{Z}}
\newcommand{\bC}{\mathbb{C}}
\newcommand{\bR}{\mathbb{R}}

\newcommand{\per}{\ell}

\numberwithin{equation}{section}
\begin{document}


\begin{titlepage}


\begin{flushright}
Imperial/TP/3-04/8\\ hep-th/0403002\\
\end{flushright}

\vskip 1cm

\begin{center}

\baselineskip=16pt

{\Large{\bf Sasaki--Einstein Metrics on $S^2\times S^3$}}

\vskip 1cm

Jerome P. Gauntlett$^{1*}$, Dario Martelli$^{2}$, James Sparks$^{2}$
and Daniel Waldram$^{2}$

\vskip 1cm

{\small{\it $^1$Perimeter Institute for Theoretical Physics\\
Waterloo, ON, N2J 2W9, Canada\\ E-mail:
jgauntlett@perimeterinstitute.ca\\}} \vskip .8cm

{\small{\it $^{2}$Blackett Laboratory, Imperial College\\  London,
SW7 2BZ, U.K.\\   E-mail: d.martelli, j.sparks,
d.waldram@imperial.ac.uk \\}}

\end{center}

\vskip 1cm

\begin{center}
\textbf{Abstract}
\end{center}

\begin{quote}
We present a countably infinite number of new explicit
co-homogeneity one Sasaki--Einstein metrics on $S^2\times S^3$, in
both the quasi-regular and irregular classes. These give rise to
new solutions of type IIB supergravity which are expected to be
dual to ${\cal N}=1$ superconformal field theories in four dimensions
with compact or non-compact $R$-symmetry and rational or irrational
central charges, respectively.
\end{quote}

\vfill \vskip 5mm
\hrule width 5.cm
\vskip 5mm

{\small {\noindent $^*$ On leave from: Blackett Laboratory, Imperial
  College, London, SW7 2BZ, U.K.\\}}

\end{titlepage}

\section{Introduction}

A Sasaki--Einstein five-manifold $X_5$ may be defined as an
Einstein manifold whose metric cone is Ricci-flat and K\"ahler --
that is, a Calabi--Yau threefold. Such manifolds provide
interesting examples of the AdS/CFT correspondence \cite{mal}. In
particular, $AdS_5\times X_5$, with suitably chosen self-dual
five-form field strength, is a supersymmetric solution of type IIB
supergravity that is conjectured to be dual to an ${\cal N}=1$ four-dimensional
superconformal field theory arising from a stack of D3-branes
sitting at the tip of the corresponding
Calabi--Yau cone~\cite{Klebanov:1998hh,Morrison:1998cs,
Figueroa-O'Farrill:1998nb,Acharya:1998db}. It is
striking that the only Sasaki--Einstein five-metrics
that are explicitly known are the round metric on $S^5$ (or
$S^5/\mathbb{Z}_3$) and the homogeneous metric $T^{1,1}$ on
$S^2\times S^3$ (or $T^{1,1}/\mathbb{Z}_2$). For $S^5$ the
Calabi--Yau cone is simply $\mathbb{C}^3$ while for $T^{1,1}$ it
is the conifold. Here we will present a countably infinite number
of explicit co-homogeneity one Sasaki--Einstein  metrics on
$S^2\times S^3$.

The new metrics were found rather indirectly. In \cite{GMSW} we
analysed general supersymmetric solutions of $D=11$ supergravity
consisting of a metric and four-form field strength, whose
geometries are the warped product of five-dimensional
anti-de-Sitter space ($AdS_5$) with a six-dimensional manifold. A
variety of different solutions were presented in explicit form. In
one class of solutions, the six-manifold is topologically $S^2
\times S^2\times T^2$. Dimensional reduction on one of the circle
directions of the torus gives a supersymmetric solution of type
IIA supergravity in $D=10$. A further T-duality on the remaining
circle of the original torus then leads to a supersymmetric
solution of type IIB supergravity in $D=10$ of the form $AdS_5\times
X_5$, with the only non-trivial fields being the metric and the
self-dual five-form. It is known~\cite{Acharya:1998db} that for
such solutions to be supersymmetric, $X_5$ should be
Sasaki--Einstein, and these are the metrics we will discuss here.
The global analysis presented in this paper, requiring $X_5$ to be
a smooth compact manifold, is then equivalent to requiring that
the four-form field strength in the $D=11$ supergravity solution
is quantised, giving a good M-theory background.

Sasaki--Einstein geometries always possess a Killing vector of
constant norm (for some general discussion see, for
example,~\cite{boyer}). If the orbits are compact, then we have a
$U(1)$ action. The quotient space is always locally
K\"ahler--Einstein with positive curvature. If the $U(1)$ action
is free then the quotient space is a K\"ahler--Einstein manifold
with positive curvature. Such Sasaki--Einstein manifolds are
called regular, and the five-dimensional compact variety are
completely classified~\cite{fried}. This follows from the fact that
the smooth four-dimensional K\"ahler--Einstein metrics with positive
curvature on the base have been classified by Tian and Yau
\cite{tian,tianyau}. These include the special cases
$\mathbb{C}P^2$ and $S^2\times S^2$, with corresponding
Sasaki--Einstein manifolds being the homogeneous manifolds $S^5$
(or $S^5/\mathbb{Z}_3$) and $T^{1,1}$ (or $T^{1,1}/\mathbb{Z}_2$),
respectively. For the remaining metrics, the base is a del Pezzo
surface obtained by blowing up $\mathbb{C}P^2$ at $k$ generic
points with $3\le k\le8$ and, although proven to exist, these
metrics are not known explicitly.

More generally, since the $U(1)$ Killing vector has constant norm,
the action it generates has finite isotropy subgroups. Only if the
isotropy subgroup of every point is trivial is the action free.
Thus in general the base -- the space of leaves of the canonical
$U(1)$ fibration --  will have orbifold singularities. This class
of metrics is called quasi-regular~\cite{boyerone}.
Recently~\cite{boyerone,Boyer:2000pg,boyerthree,boyerfour}, a rich
set of examples of quasi-regular Sasaki--Einstein metrics have
been shown to exist on $\#l(S^2\times S^3)$ with $l=1,\dots,9$,
but they have not yet been given in explicit form\footnote{Very
recently an infinite class of explicit inhomogeneous Einstein
metrics have been constructed in \cite{Hashimoto:2004kc}. These
include Einstein metrics on $S^2\times S^3$, but they are not
expected to be Sasaki--Einstein.}. In particular, there are 14
inhomogeneous Sasaki--Einstein metrics on $S^2\times S^3$.

Some of the new metrics presented here are foliated by $U(1)$ and
are thus in the quasi-regular class. We do not yet know if they
include any of the 14 discussed in \cite{Boyer:2000pg}. We also
find Sasaki--Einstein metrics where the Killing vector has
non-compact orbits, which are hence {\em irregular}. These seem to be
the first examples of such metrics.

It is worth emphasising that the isometries generated by the
canonical Killing vector on the Sasaki--Einstein manifold are dual
to the $R$-symmetry in the four-dimensional superconformal field
theory. Thus the regular and quasi-regular examples are dual to
field theories with compact $U(1)$ $R$-symmetry. In contrast, the
irregular examples are dual to field theories with a non-compact
$R$-symmetry. In other words, these theories would
be invariant under the superconformal algebra but not the
superconformal group. We will also calculate the volumes of the
new Sasaki--Einstein metrics. In the dual conformal field theory
these are inversely related to the central charge of the conformal
field theory. We show that the metrics associated with compact
$U(1)$ $R$-symmetry are associated with rational central
charges while those with non-compact $R$-symmetry are associated with
irrational central charges.

\section{The metrics}
\label{metric}

Our starting point is the explicit local metric given by the line
element~\cite{GMSW}:
\begin{equation}
\label{tinky}
\begin{aligned}
   ds^2 &= \frac{1-cy}{6}(d\theta^2+\sin^2\theta
      d\phi^2)+\frac{1}{w(y)q(y)}
      dy^2+\frac{q(y)}{9}[d\psi-\cos\theta d\phi]^2 \\ & \qquad \qquad
      + {w(y)}\left[d\alpha +\frac{ac-2y+y^2c}{6(a-y^2)}
         [d\psi-\cos\theta d\phi]\right]^2
\end{aligned}
\end{equation}
with
\bea
 w(y) &=& \frac{2(a-y^2)}{1-cy}\nn
 q(y) &=& \frac{a-3y^2+2cy^3}{a-y^2}~.
\eea
A direct calculation shows that this metric is Einstein with
$\mathrm{Ric}=4 g$, for all values of the constants $a,c$.
Locally the space is also Sasaki.
Note that the definition of Sasaki--Einstein
can be given in several (more-or-less) equivalent ways. For
example, one can define a Sasaki--Einstein geometry in terms of
the existence of a certain contact structure, or in terms of the
existence of a solution to the Killing spinor equation.
The simplest way to demonstrate that we have a local Sasaki-Einstein
metric is to write the metric in a canonical way, which we do in
section~\ref{sasaki}. The corresponding local Killing spinors
are then easily obtained (see the discussion in \cite{Gibbons:2002th}).

The next step is to analyse when we can extend the local
expression for the metric to a global metric on a complete
manifold. In this section we will demonstrate that this can indeed
be done, with the manifold being $S^2\times S^3$. The point of
section~\ref{sasaki} will be to show that the Sasaki structure
also extends globally, ensuring we have globally defined Killing
spinors.

We find the global extensions in two steps. First we show that,
for given range of $a$, one can choose the ranges of the
coordinates $(\theta,\phi,y,\psi)$ so that this ``base space"
$B_4$ (forgetting the $\alpha$ direction) is topologically the
product space $S^2\times S^2$. The second step is to show that,
for a countably infinite number of values of $a$ in this range,
one can choose the period of $\alpha$ so that the five-dimensional
space is then the total space of an $S^1$ fibration (with $S^1$
coordinate $\alpha$) over $B_4$. Topologically this five-manifold
turns out to be $S^2\times S^3$. A detailed analysis of the
possible values for $a$ is the content of section~\ref{solutions}.

As we see in section~\ref{special}, when $c=0$ the
metric is the local form of the standard homogeneous metric on
$T^{1,1}$ (the parameter $a$ can always be rescaled by a
coordinate transformation). Thus we will focus on the case $c\ne
0$. In this case we can, and will, rescale $y$ to set $c=1$. This
leaves us with a local one-parameter family of metrics,
parametrised by $a$.

\subsubsection*{The base $B_4$}

First, we choose $0\le\theta\le \pi$ and $0\le \phi \le 2\pi$ so
that the first two terms in~\p{tinky}, at fixed $y$, give the
metric on a round two-sphere. Moreover, the two-dimensional
$(y,\psi)$-space, defined by fixing $\theta$ and $\phi$, is clearly
fibred over this two-sphere. To analyse the fibre, we fix the range of
$y$ so that $1-y>0$, $a-y^2>0$, which implies that $w(y)>0$. We
also demand that $q(y) \ge 0$ and that $y$ lies between two zeroes
of $q(y)$, {\it i.e.} $y_1\le y\le y_2$ with $q(y_i)=0$. Since the
denominator of $q(y)$ is always positive, $y_i$ are roots of the
cubic
\be\label{thecubic}
 a-3y^2+2 y^3=0~.
\ee
All of these conditions can be met if we fix the range of $a$ to
be \be 0<a< 1~. \ee In particular, for this range, the cubic has
three real roots, one negative and two positive, and $q(y)\ge 0$
if $y_1$ is taken to be the negative root and $y_2$ the smallest
positive root. The case $a=1$ is special since the two positive
roots coalesce into a single double root at $y=1$. In fact we will
show in section 5
that when $a=1$ the metric \p{tinky} is locally that of $S^5$.

We now argue that by taking $\psi$ to be periodic with period
$2\pi$, the $(y,\psi)$-fibre, at fixed $\theta$ and $\phi$, is
topologically a two-sphere. To see this, note the space is a
circle fibred over an interval, $y_1\le y\le y_2$, with the circle
shrinking to zero size at the endpoints, $y_i$. This will be
diffeomorphic to a two-sphere provided that the space is free from
conical singularities at the end-points. For $0<a<1$, near either
root we have $q(y)\approx q'(y_i)(y-y_i)$. Fixing $\theta$ and $\phi$
in~\p{tinky} (and ignoring the $\alpha$ direction for the time being),
this gives the metric near the poles $y=y_i$ of the two-sphere
\be
\frac{1}{w(y_i)q'(y_i)(y-y_i)}dy^2+\frac{q'(y_i)(y-y_i)}{9}d\psi^2~.
\ee
Introducing the co-ordinate $R=[4(y-y_i)/w(y_i)q'(y_i)]^{1/2}$ this
can be written as
\be dR^2+\frac{q'(y_i)^2w(y_i)R^2}{36}d\psi^2~. \ee
We now note the remarkable fact that since $q'(y_i)=-3/y_i$ and
$w(y_i)=4y_i^2$, it follows that $q'(y_i)^2w(y_i)/36=1$ at any root of
the cubic. Thus the potential conical singularities at the poles
$y=y_1$ and $y=y_2$ can be avoided by choosing the period of $\psi$ to
be $2\pi$.

Note that the properties of the function $q(y)$ allow the
introduction of an angle $\zeta(y)$ defined by
\bea\label{coszeta}
\cos\zeta&=&
q(y)^{1/2}=\left(\frac{a-3y^2+2y^3}{a-y^2}\right)^{1/2}\nn
\sin\zeta&=& -\frac{2y}{w(y)^{1/2}}
\eea
with $\zeta$ ranging from $\pi/2$ to $-\pi/2$ between the two
roots (\textit{cf.}~\cite{GMSW}).
It is not simple to explicitly change coordinates from $y$
to $\zeta$, so we continue to work with $y$.

At fixed value of $y$ between the two roots, the
$(\psi,\theta,\phi)$-space is a $U(1)=S^1$ bundle, parametrised by
$\psi$, over the round two-sphere parametrised by $\theta$ and
$\phi$. Such bundles are, up to isomorphism, in one-to-one
correspondence with $H^2(S^2;\mathbb{Z})=\mathbb{Z}$, which is the
Chern number, or equivalently the integral of the curvature
two-form over the base space. In our case, this integral is
\be \frac{1}{2\pi}\int_{S^2} d(-\cos\theta d\phi) = 2~.\ee
This identifies the three-space at fixed $y$ as the Lens space
$S^3/\mathbb{Z}_2 = \mathbb{R}P^3$. Equivalently, it is the total
space of the bundle of unit tangent vectors of $S^2$. This
establishes that the four-dimensional base manifold is a
two-sphere bundle over a two-sphere. Moreover, the
$\mathbb{R}^2=\mathbb{C}$ bundle over $S^2$ obtained by deleting
the north pole from each two-sphere fibre is just the tangent
bundle of $S^2$.

In general, oriented $S^2$ bundles over $S^2$ are classified up to
isomorphism by an element of $\pi_1(SO(3))=\mathbb{Z}_2$. One
constructs any such bundle by taking trivial bundles, {\it i.e.}
products, over the northern and southern hemispheres and gluing
them together along the equator, with the appropriate group
element. The gluing is given by a map from the equatorial $S^1$
into $SO(3)$. The topology of the bundle depends only on the
homotopy type of the map and hence there are only two bundles,
classified by an element in $\pi_1(SO(3))=\mathbb{Z}_2$, one
trivial and one non-trivial\footnote{The non-trivial bundle is
obtained by adding a
  point to the fibres of the chiral spin bundle of $S^2$ and is not a
  spin manifold. It gives the same manifold as the Page instanton
  \cite{page} on $\mathbb{C}P^2\# \overline{\mathbb{C}P}^2$.}.
In fact in our case we have the trivial fibration and hence,
topologically, the four-dimensional base space is simply $S^2\times
S^2$ as claimed earlier.

To see this, consider the gluing element of $\pi_1(SO(3))$
corresponding to the map
\be \phi\rightarrow\left(\begin{array}{ccc}\cos(N\phi) &
-\sin(N\phi) & 0 \\ \sin(N\phi) & \cos(N\phi) & 0
\\ 0 & 0 & 1\end{array}\right)\label{glue}\ee
where $\phi$ is a coordinate on the equator of the base $S^2$,
with $0\leq \phi \leq 2\pi$. Here, the upper-left $2\times2$ block
of the matrix describes the twisting of the equatorial plane in
the fibre (the $\psi$ coordinate in our metric) and the
bottom-right entry refers to the ``polar'' direction. If one
projects out the polar direction, leaving the $2\times 2$ block in
the upper-left corner, the above map then gives the element $N\in
\pi_1(U(1)) = \mathbb{Z}$ corresponding to the Chern class of a
$U(1)=SO(2)$ bundle over $S^2$ -- this bundle is just a charge
$N$ Abelian monopole.

Now, the above $SO(3)$ matrix is well-known to correspond (with a
choice of sign) to the $SU(2)$ matrix
\be \left(\begin{array}{cc}e^{iN\phi/2} & 0 \\
0 & e^{-iN\phi/2}\end{array}\right)\ee
where we recall that $SU(2)$ is the simply-connected double-cover
of $SO(3)$. It follows that, for $N$ even, the resulting curve in
$SU(2)=S^3$ is a closed cycle, and therefore corresponds to the
trivial element of $\pi_1(SO(3)) = H_1(SO(3);\mathbb{Z}) =
\mathbb{Z}_2$. Thus all even $N$ give topologically product spaces
$S^2\times S^2$. For $N$ odd, the curve in $SU(2)$ is not closed
-- it starts at one pole of $S^3$ and finishes at the antipodal
pole. Of course, when projected to $SO(3)$ this curve now becomes
closed and thus represents a non-trivial cycle. In fact this is
the generator of $\pi_1(SO(3))$. Thus all the odd $N$ give the
non-trivial $S^2$ bundle over $S^2$. In particular, recall that
for us the $U(1)$ bundle corresponding to $\psi$ was $\bR P^3$
with $N=2$, and thus we have topologically a product space
\begin{equation}
   B_4 = S^2 \times S^2~.
\end{equation}

In what follows it will be useful for us to have an explicit basis
for the homology group
$H_2(B_4;\mathbb{Z})=\mathbb{Z}\oplus\mathbb{Z}$ of
two-dimensional cycles on $B_4=S^2\times S^2$. The natural choice
is simply the two $S^2$ cycles $C_1$, $C_2$ themselves, but since
our metric on $B_4$ is not a product metric, it is not immediately
clear where these two two-spheres are. In fact, we can take $C_1$
to be the fibre $S^2$ at some fixed value of $\theta$ and $\phi$
on the round $S^2$. Returning to the metric~\p{tinky} we note that
there are two other natural copies of $S^2$ located at the south
and north poles of the fibre, {\it i.e.} at the roots $y=y_1$ and
$y_2$, or equivalently $\zeta=\pi/2$ and $\zeta=-\pi/2$. Call
these $S_1$ and $S_2$. Then the other cycle\footnote{One way to
check this is to work out the intersection numbers of the cycles.
We have $C_1^2=C_2^2=0$, $C_1\cdot C_2=1$, $S_1^2=2$, $S_2^2=-2$,
$S_1\cdot S_2=0$, $S_1\cdot C_1=S_2\cdot C_1=1$.} is just $C_2 =
S_2 + C_1 = S_1 - C_1$. Thus we have
\begin{equation} \label{CSrel}
   2C_1 = S_1 - S_2 \qquad
   2C_2 = S_1 + S_2~.
\end{equation}
For completeness let us also give explicitly the dual elements in the
cohomology $H^2(S^2\times S^2;\bZ)$. We have
\begin{equation}
\label{cohom}
\begin{aligned}
   \omega_1 &= \frac{1}{4\pi}\cos\zeta
         d\zeta \wedge (d\psi-\cos\theta d\phi)
      + \frac{1}{4\pi}\sin\zeta\sin\theta d\theta \wedge d\phi \\
   \omega_2 &= \frac{1}{4\pi}\sin\theta d\theta \wedge d\phi
\end{aligned}
\end{equation}
with $\int_{C_i}\omega_j=\delta_{ij}$.

\subsubsection*{The circle fibration}

Now we turn to the fibre direction, $\alpha$, of the full
five-dimensional space. It is convenient to write the
five-dimensional metric~\p{tinky} in the form
\be ds^2=ds^2(B_4)+w(y)(d\alpha+A)^2 \ee
where $ds^2(B_4)$ is the non-trivial metric on $S^2\times S^2$ just
described, and the local one-form $A$ is given by
\be A= \frac{a-2y+y^2}{6(a-y^2)}[d\psi-\cos\theta d\phi]~. \ee
%
Note that the norm-squared of the Killing vector
$\partial/\partial \alpha$ is $w(y)$, which is nowhere-vanishing.

In order to get a compact manifold, we would like the $\alpha$
coordinate to describe an $S^1$ bundle over $B_4$. We take
\be
0\leq\alpha\leq2\pi\per
\ee
where the period $2\pi\per$ of $\alpha$
is \textit{a priori} arbitrary. Thus,
rescaling by $\per^{-1}$, we have that $\per^{-1}A$ should be a
connection on a $U(1)$ bundle over $B_4=S^2\times S^2$. However,
this puts constraints on $A$. In general, such $U(1)$ bundles are
completely specified topologically by the gluing on the equators
of the two $S^2$ cycles, $C_1$ and $C_2$. These are measured by
the corresponding Chern numbers in $H^2(S^2;\bZ)=\bZ$ which we
label $p$ and $q$. The corresponding five-dimensional spaces will
be denoted by $Y^{p,q}$. In general, we will find that for any $p$
and $q$, such that $0<q/p<1$, we can always choose $\per$ and the
parameter $0<a<1$ such that $\per^{-1}A$ is a \textit{bona fide}
$U(1)$ connection. For $p$ and $q$ relatively prime, which we can
always achieve by taking an appropriate $\per$, {\it i.e.} the
maximal period, it turns out that $Y^{p,q}$ are all topologically
$S^2\times S^3$. We now fill in the details of this argument.

The essential point is to show that $\per^{-1}A$ is a connection
on a $U(1)$ bundle. As mentioned above, a $U(1)$ bundle over
$B_4=S^2\times S^2$ is characterised by the two Chern numbers $p$
and $q$. These are given by the integrals of the $U(1)$-curvature
two-form $\per^{-1}dA/2\pi$ over the two-cycles $C_1$
and $C_2$ which form the basis of $H_2(S^2\times
S^2;\bZ)=\bZ\oplus\bZ$. Let us define the two periods $P_1$ and
$P_2$ as
\be
   P_i=\frac{1}{2\pi}\int_{C_i} dA
\ee
which, in general, will be functions of $a$. The corresponding
integrals of $\per^{-1}dA/2\pi$ must give the Chern numbers $p$
and $q$. That is, we require $P_1=\per p$ and $P_2=\per q$. Since
we are free to choose $\per$, the only constraint is that we must
find $a$ such that
\begin{equation}
\label{ratcond}
   P_1/P_2 = p/q~.
\end{equation}
In particular, we will choose $\per$ so that $p$ and $q$ are
coprime. Then with
\be\label{defperiodal}
\per=P_1/p=P_2/q
\ee
we get a five-dimensional
manifold which is an $S^1$ bundle over $B_4=S^2 \times S^2$ with
winding numbers $p$ and $q$, denoted by $Y^{p,q}$. We will find
that~\eqref{ratcond} can be satisfied for a countably infinite
number of values of $a$.

Let us first check that $dA$ is properly globally defined. Recall
that, in general, a connection one-form is not a globally
well-defined one-form -- if it is globally well-defined, the
curvature is exact and the bundle is topologically trivial.
Rather, a connection one-form is defined only locally in patches,
with gauge transformations between the patches. However, the
curvature is a globally well-defined smooth two-form. Let us now
check this is true for our metric. At fixed value of $y$ between
the two roots, $y_1<y<y_2$, we see that $A$ is proportional to the
``global angular form" on the $U(1)$ bundle with fibre $\psi$ and
base parametrised by $(\theta,\phi)$. This is actually globally
well-defined since a gauge transformation on $\psi$ is cancelled
by the corresponding gauge transformation of the connection
$-\cos\theta d\phi$. Thus in fact $dA$ is exact on a slice of the
four-manifold $B_4$ at fixed $y$. This is also clearly true on the
whole of the total space of the ``cylinder bundle" obtained by
deleting the north and south poles of the fibres of the $S^2$
bundle. One must now check that $dA$ is smooth as one approaches
the poles of the fibres. There are two terms, and the only term of
concern smoothly approaches a form proportional to $dy\wedge
d\psi$ near the poles, at fixed value of $\theta$ and $\phi$. We
must now recall that the proper radial coordinate is ({\it e.g.} at
the south pole) $R\propto (y-y_i)^{1/2}$. Thus $dy\propto RdR$,
and this piece of $dA$ is a smooth function times the canonical
volume form $RdR\wedge d\psi$ on the open subset of $\mathbb{R}^2$
in the fibre near to the poles. Thus $dA$ is a globally-defined
smooth two-form on $B_4 = S^2 \times S^2$, and thus represents an
element of $H^2_{\mathrm{de\ Rham}}(B_4)$.

To calculate the periods $P_i$ it is easiest to first calculate the
integrals of $dA$ over the cycles $S_i$ at the north and
south poles of the $(y,\psi)$ fibre. We find
\be
   \frac{1}{2\pi}\int_{S_i}dA =
      \frac{1}{2\pi}\int_{S_i}\frac{a-2y_i+y_i^2}{6(a-y_i^2)}
      \sin\theta d\theta d\phi=\frac{y_i-1}{3y_i}
\ee
and hence, given the relations~\eqref{CSrel}, we have
\bea
\label{ratperiods}
   P_1&=&\frac{y_1-y_2}{6 y_1 y_2}\nn
   P_2&=&\frac{2 y_1 y_2-y_1-y_2}{6 y_1 y_2}
   =-\frac{(y_1-y_2)^2}{9y_1y_2}
\eea
and so
\be
   \frac{P_1}{P_2}=\frac{3}{2(y_2-y_1)}~.
\ee
Thus our requirement~\eqref{ratcond} is that
\begin{equation}
   \text{$y_2-y_1$ is rational.}
\end{equation}

In the next section we will show that there are infinitely many values
of $a$ for which this is true, in the range $0<a<1$ and with
$0<q/p<1$. Furthermore, as shown in Appendix A, for any $p$ and $q$
coprime, the space $Y^{p,q}$ is topologically $S^2\times S^3$.
This then completes our argument
about the regularity of the metrics on $S^2\times S^3$.

Our new metrics admit a number of Killing vectors which generate
isometries. Clearly there is a $U(1)$ generated by
$\partial/\partial\alpha$. In addition there is an $SU(2)\times
U(1)$ action. To see this, we write the metric in terms of
left-invariant one-forms $\sigma_i$, $i=1,2,3$ on $SU(2)$
\begin{equation}
\label{tinkytwo}
\begin{aligned}
   ds^2&= \frac{1-y}{6}(\sigma_1^2+\sigma_2^2)
     +\frac{1}{w(y)q(y)}dy^2
     +\frac{q(y)}{9}\sigma_3^2 \\ & \hspace{3.5cm}
     + {w(y)}\left[d\alpha
        +\frac{a-2y+y^2}{6(a-y^2)} \sigma_3\right]^2~.
\end{aligned}
\end{equation}
This displays the fact that the metrics admit an $SU(2)$
left-action and a $U(1)$ right-action. Together with the $U(1)$
isometry generated by $\partial/\partial\alpha$, this gives an
isometry group $SU(2)\times_{\bZ_2} U(1)^2$, where we have noted
that the element $(-1_2,-1,-1)$ acts trivally.

We end by noting that the volume of these spaces is given by
\be
   \mathrm{vol} = \frac{4\pi^3}{9}\per\;(y_1-y_2)(y_1+y_2-2)~.
\ee

\section{All solutions for the parameter $a$}
\label{solutions}

We have shown that it is necessary and sufficient that $P_1/P_2 =
p/q$ is rational in order to get metrics on a complete manifold.
Clearly it is
sufficient that the roots $y_1$ and $y_2$ of the cubic (\ref{thecubic})
(and hence all three of the roots) are rational. This leads to a number
theoretic analysis which is presented below, and we find an
infinite number of values of $a$ for which the roots are rational.
In these cases the volume of the manifolds are rationally related
to the volume of the round five-sphere. We will argue later that
these values of $a$ give rise to quasi-regular Sasaki--Einstein
manifolds. However, it is also possible to achieve rational
$P_1/P_2$ for an infinite number of values of $a$ when the roots
are \emph{irrational} as the following general analysis reveals.
We will see later that these cases give rise to irregular
Sasaki--Einstein metrics.

\subsubsection*{General case}

Assume  $P_1/P_2 = p/q$ is rational.  As discussed above, this implies
that
\bea y_2-y_1 = \frac{3q}{2p} \equiv \lambda~.\eea
We first observe that the three roots of the cubic satisfy: \bea
y_1+y_2+y_3&=&3/2\nn y_1y_2+y_1y_3+y_2y_3&=&0\nn 2y_1y_2y_3&=&-a~.
\eea Next note that $y_1+y_2=0$ if and only if $a=0$, which is
excluded from our considerations. Then we deduce that
\be y_1+y_2=\frac{2}{3}(y_1^2+y_1y_2+y_2^2)\ee
with the third root given by $y_3=\frac{3}{2}-y_1-y_2$.
One may now solve for $y_1$ in terms of $\lambda$.
Since $y_1$ is required to be the smallest root of the
cubic, one takes the smaller of the two roots to obtain
\be y_1 =
\frac{1}{2}\left(1-\lambda-\sqrt{1-\lambda^2/3}\right)~.\label{nobby}\ee
Since $y_2>y_1$ this means that $0<\lambda\leq \sqrt{3}$, where
the upper bound ensures that $y_1$ is real. Notice that $y_1<0$,
as it should be.

We now require that $y_1$ be a root of the cubic. To ensure this,
we simply define
\be\label{aintermsofy1} a = a(\lambda) = 3y_1^2(\lambda) -
2y_1^3(\lambda)~.\ee
One requires that $0<a<1$ which is in fact automatic for the range
of $\lambda$ already chosen. However, some values of $a$ are
covered twice since the function $a(\lambda)$ has a maximum of 1
at the value $\lambda=3/2$. Specifically, this range is easily
computed to be $1/2\leq a \leq 1$. The range $0<a<1/2$ is covered
only once. Finally, we must ensure that $y_2 = y_1 +\lambda$ is the smallest
positive root. Comparing with  $y_3=\frac{3}{2}-y_1-y_2$, we see that
this implies that $\lambda<\frac{3}{2}$.

To summarise, any rational value of $\lambda=3q/2p$ is allowed
within the range $0<\lambda< 3/2$, and this is achieved by
choosing $a$ given by \p{nobby}, \p{aintermsofy1}. Moreover, the
range of $a$ is $0<a< 1$ and is covered in a monotonic increasing
fashion. The period of the coordinate $\alpha$ is given by
$2\pi\per$ where
\begin{equation}
   \label{per-pq}
   \per = \frac{q}{3q^2-2p^2+p(4p^2-3q^2)^{1/2}}~.
\end{equation}
Note that we can also recast our formula for the volume
in terms of $p$ and $q$ to give
\be
\label{volume}
   \mathrm{vol} =
     \frac{q^2[2p+(4p^2-3q^2)^{1/2}]}{3p^2[3q^2-2p^2+p(4p^2-3q^2)^{1/2}]}
     \pi^3
\ee
which is generically an irrational fraction of the volume $\pi^3$
of a unit round $S^5$. This implies that we have irrational
central charges in the dual superconformal field theory. In
fact the result is stronger: the central charge can be written only in
terms of square-roots of rational numbers. Note that by setting $q=1$ and
letting $p$ become large, we see that the volume can be arbitrarily
small. The largest volume given by our metrics on $Y^{p,q}$ occurs for
$p=2,q=1$ with $\mathrm{vol}\approx0.29\pi^3$, and corresponds to an
irrational case.

\subsubsection*{Case of rational roots}

We now show that for a countably infinite number of values of $a$
the roots $y_1$ and $y_2$ are rational. First note
from~(\ref{nobby}) that $y_1$ is rational if and only if
$1-\lambda^2/3$ is the square of a rational. Then $y_2$ (and also
$y_3$) is rational since $\lambda$ is necessarily rational.
Substituting $\lambda=3q/2p$, the problem reduces to finding all
solutions to the quadratic diophantine
\bea
4p^2 - 3q^2 = n^2\label{dio}
\eea
where $p,q\in \mathbb{N}$, $n\in \mathbb{Z}$, $(p,q)=1$, $q<p$.

To find the general solution, we proceed as follows. First define
$r=2p$, so that \p{dio} may be written \bea (r-n)(r+n)=3q^2~. \eea
We first argue that a prime factor $t>3$ of $r+n$ must appear with
an even power. Suppose it did not. We have $t$ divides $q^2$ and
so $t$ divides $q$. Then, since the power of $t$ in $q^2$ is
clearly even, $t$ must divide $r-n$. This is now a contradiction
since $t$ now divides both $p$ and $q$ which are, by assumption,
coprime. Using a similar argument for $t=3$ we conclude that
\bea r+n&=&3A^2 2^{k_1}\nn r-n&=&B^22^{k_2} \eea where $A,B,
k_i\in \mathbb{N}$ and we have used the freedom of switching the
sign of $n$ in fixing the factor of 3. Moreover, $A,B$ satisfy
\be\label{condsAB} (A,B)=1,\quad B\neq 0 \mod 3,\quad A,B\neq 0
\mod 2 ~.\ee We deal with the factors of 2 in two steps. First, if
$n$ is odd then $k_1=k_2=0$ and we have the following solutions
\bea p = \frac{1}{4}(3A^2+B^2), \quad q = AB, \eea with
$n=\frac{1}{2}(3A^2-B^2)$. To ensure that $p>q$ we need to also
impose \be A>B,\quad {\rm or}\quad B>3A~. \ee Alternatively, for
solutions with $n$ even, one can show that $q$ is always even and
$p$ is then always odd. A little effort reveals two families of
solutions. The first is: \bea p=3A^2+B^2 2^{2k}, \quad q=AB2^{k+2}
\eea with $n=2(3A^2-B^22^{2k})$, $k\ge 1$ and $A,B$ satisfying
\p{condsAB} and also \be A>B2^k,\quad {\rm or} \quad B2^k>3A ~.\ee
The second is: \bea p=3A^22^{2k}+B^2, \quad q=AB2^{k+2} \eea with
$n=2(3A^22^{2k}-B^2)$, $k\ge 1$ and $A,B$ satisfying \p{condsAB}
and also \be A2^k>B,\quad {\rm or} \quad B>3A2^k ~.\ee

Finally, returning to the volume formula~\eqref{volume}, we note that
when the roots are rational, since $4p^2-3q^2=n^2$, clearly the volume
is a rational fraction of that of the unit round $S^5$. This corresponds to
rational central charges in the dual superconformal field theory.

\section{The metrics in canonical form}
\label{sasaki}

At this stage we have shown that we have a countably infinite
number of new Einstein metrics on $S^2\times S^3$.
We now establish that the geometries admit a Sasaki structure which
extends globally and that the metrics admit globally defined Killing spinors.
We do this by first showing that the metric can be written in a
canonical form implying it has a local Sasaki structure.
We then show that this Sasaki-Einstein structure, defined in terms
of contact structures, is globally well-defined.
Finally, given we have a simply-connected five-dimensional
manifold with a spin structure, using theorem~3 of~\cite{fried},
we see that this implies we have global Killing spinors.

Employing the change of coordinates $\alpha=-\beta/6-c\psi'/6$,
$\psi=\psi'$ the local metric \p{tinky} becomes
\begin{equation}
\label{canmetric}
\begin{aligned}
   ds^2 &=  \frac{1-cy}{6}(d\theta^2+\sin^2\theta d\phi^2)
      + \frac{dy^2}{w(y)q(y)}
      + \frac{1}{36} w(y)q(y)(d\beta+c\cos\theta d\phi)^2 \\
      &\qquad\qquad
      + \frac{1}{9}[ d\psi'-\cos\theta d\phi
         +y(d\beta+c\cos\theta d\phi)]^2
\end{aligned}
\end{equation}
%
where we have temporarily reinstated the constant $c$. This has
the standard form
\bea
\label{SE}
   ds^2= ds^2_4+ \left(\tfrac{1}{3}d\psi' + \sigma\right)^2
\eea
where $ds^2_4$ is a local K\"ahler--Einstein metric and the form
$\sigma$ satisfies $d\sigma=2 J_4$. We have the local K\"ahler
form
\begin{equation}
   J_4 = \frac{1}{6}(1-cy) \sin\theta\,d\theta \wedge d\phi
      + \frac{1}{6}dy \wedge (d\beta+c\cos\theta d\phi)~.
\end{equation}
In the quasi-regular case, $J_4$ extends globally to the K\"ahler
form on the four-dimensional base orbifold.

One can check explicitly that the four-dimensional metric is
K\"ahler--Einstein and has Ricci-form equal to six times the K\"ahler
form. This form of the metric~\eqref{SE} is the standard one for a locally
Sasaki--Einstein metric, with $\partial/\partial\psi'$ the
constant norm Killing vector. As noted earlier, the normalisation
is canonical with the Ricci tensor being four times the metric.

Note that the $SU(2)\times U(1)^2$ isometry group of our
metrics~\eqref{tinkytwo} implies that we can introduce a set of
$SU(2)$ left-invariant one-forms, $\tilde\sigma_i$, and write $ds^2_4$
in~\eqref{SE} locally as a bi-axially squashed $SU(2)$-invariant
K\"ahler--Einstein metric. The most general metric of this type was
found in~\cite{Gibbons-Pope}, where it was shown to depend on two
parameters (one being the overall scale). Global properties of such
metrics were then discussed, to some extent, in~\cite{Pedersen}. To
recast our metric in the form given
in~\cite{Gibbons-Pope}, we introduce $\rho^2 = 2(1 - y)/3$
and $\tilde \sigma_i$ to get, with $c=1$,
\bea
   ds^2_4&=&\frac{1}{\Delta}d\rho^2
      +\frac{\rho^2}{4}\left(\tilde\sigma_1^2+\tilde\sigma_2^2
         +\Delta\tilde\sigma_3^2\right)\nn
   \Delta&=&1+\frac{4(a-1)}{27}\frac{1}{\rho^4}-\rho^2~.
\eea

We would now like to confirm that the Sasaki structure extends globally.
To do so, we will work with the definition of
a Sasaki--Einstein manifold given in terms of a contact structure.
Given the metric extends globally, this is equivalent to the
condition that the Killing vector $\de/\de\psi'$ and hence the
dual one-form $\frac{1}{3}d\psi'+\sigma$ with $d\sigma=2J_4$
in~\eqref{SE} are globally defined.
To show this we simply consider these objects in
the original coordinates~\eqref{tinky}. We first observe that the
Killing vector field of the Sasaki structure is given by
\be\label{sasvec} \frac{\partial}{\partial
\psi'}=\frac{\partial}{\partial \psi}-\frac{1}{6}
\frac{\partial}{\partial \alpha}
\ee
which is globally well defined, since both of the vectors on the right
hand side are globally well defined. Since this vector has constant norm,
the dual one-form
\be\label{dualonef} [ d\psi'-\cos\theta
d\phi+y(d\beta+c\cos\theta d\phi)] \ee
must also be globally well defined. It is interesting to see this in the
original coordinates. This one-form can be written as the sum of two
one-forms:
\bea\label{oneformd}
-6y\left[d\alpha+\frac{ac-2y+y^2c}{6(a-y^2)} (d\psi-\cos\theta
   d\phi)\right]\nn
+ \frac{a-3y^2+2cy^3}{a-y^2}\left(d\psi-\cos\theta d\phi\right)~.
\eea
From the analysis of section 2, we conclude that the first
one-form in this expression is globally well-defined since it is,
up to a smooth function, just the dual of the globally defined
vector $\partial/\partial\alpha$. This form is also the so-called
global angular form on the total space of the $U(1)$ bundle
$U(1)\hookrightarrow Y^{p,q}\rightarrow S^2\times S^2$. Now, the
one-form $\left(d\psi-\cos\theta d\phi\right)$ is a global angular
form on the $U(1)$ bundle $U(1)\hookrightarrow
\mathbb{R}P^3\rightarrow S^2$ at fixed $y$ between the two
roots/poles. Although this one-form is not well defined at the
poles $y=y_i$, the second one-form in \p{oneformd} \emph{is}
globally defined, since the pre-factor vanishes smoothly at the
poles. Finally, the exterior derivative of \p{dualonef} is also
clearly well defined and is equal to $6J_4$. Thus we conclude that
the Sasaki structure is globally well defined and that we have a
countably infinite number of Sasaki--Einstein manifolds.

Recall that in the original co-ordinates we argued that $\psi$ was periodic
with period $2\pi$ and that $\alpha$ was periodic with period
$2\pi\per=2\pi P_1/p=2\pi P_2/q$. The form of \p{sasvec} shows that,
for general values of $a$ discussed in section 3, the orbits of
the vector $\partial/\partial\psi'$ are not closed. In this case
the Sasaki--Einstein metric is irregular. There is no
K\"ahler--Einstein ``base manifold'' or even ``base orbifold'' for
this case. Since the orbits of $\partial/\partial\psi'$ are dense
in the torus defined by the Killing vectors
$\partial/\partial\psi$ and  $\partial/\partial\alpha$, these
Sasaki--Einstein metrics are, more precisely, irregular of rank
two.

However, in the special case that
\be\label{blah}
\frac{6P_1}{p} = \frac{6P_2}{q}=\frac{s}{r}
\ee
for suitably chosen integers $r,s$, we can choose
$\psi'$ to be periodic with period $ 2\pi s$.
This case occurs only when the roots of the cubic are rational.
The Sasaki--Einstein metrics are then quasi-regular. Indeed, if
the base were a manifold, it would have to be in Tian and Yau's
list. Notice also that our metrics admit an $SU(2)$ action. As
discussed in \cite{Dancer}, these two facts are mutually
exclusive, except in the cases $\mathbb{C}P^2$ and
$\mathbb{C}P^1\times \mathbb{C}P^1$ with canonical metrics. Thus
we must be in the quasi-regular class. The base space is then an
orbifold -- it would be interesting to better understand their
geometry.

\section{Special cases}
\label{special}

We discuss here the two special cases that were mentioned earlier.

\subsection*{Case 1: $T^{1,1}$}

First consider $c=0$, and then rescale to set $a=3$. Starting with
\p{tinky} and introducing the coordinates $\cos\omega =
y$, $\nu=6\alpha$ we obtain
\bea ds^2=\frac{1}{6}(d\theta^2+\sin^2\theta
+d\omega^2+\sin^2\omega d\nu^2)+ \frac{1}{9}[d\psi-\cos\theta
d\phi-\cos\omega d\nu]^2~.
\eea
If the period of $\nu$ is $2\pi$ we see that the four-dimensional
base metric orthogonal to $\partial_\psi$ is now the canonical
metric on $S^2\times S^2$, and if we choose the period of $\psi$
to be $4\pi$ we recover the metric on $T^{1,1}$. If the period is
taken to be $2\pi$, as in the rest of the family with $c\ne 0$, we
get $T^{1,1}/\mathbb{Z}_2$. Both of these metrics are well known
to be Sasaki--Einstein.

\subsection*{Case 2: $S^5$}

Next, returning to $c=1$ and setting $a=1$, we introduce the new
coordinates $1-y=\frac{3}{2}\sin^2\sigma$, $\psi=\psi''-\beta$ and
$\alpha=-\psi''/6$ in~\p{tinky} to get
\begin{equation}
\begin{aligned}
   ds^2 &=
      d\sigma^2 + \frac{1}{4}\sin^2\sigma(d\theta^2+\sin^2\theta
      d\phi^2)+\frac{1}{4}\cos^2\sigma\sin^2\sigma(d\beta+\cos\theta
      d\phi)^2 \\ &\qquad
      + \frac{1}{9}\left[
         d\psi''-\frac{3}{2}\sin^2\sigma(d\beta+\cos\theta d\phi)
         \right]^2~.
\end{aligned}
\end{equation}
If the period of $\beta$ is
taken to be $4\pi$ the base metric is the Fubini-Study metric on
$\mathbb{C}P^2$. If the period of $\psi''$ is taken to be $6\pi$
we get the round metric on $S^5$. If the period of $\psi''$ is
taken to be $2\pi$ we obtain the Lens space $S^5/\mathbb{Z}_3$.
Both of these metrics are also well known to be Sasaki--Einstein.

We can view this case as a limit $a=1$ of our family of solutions.
In this case $\psi$ has period $2\pi$ implying $\beta$ has period
$2\pi$. Since $y_1=-1/2$, $y_2=1$ we take $P_1=P_2=1/2$ and
$p=q=1$ and hence the period $\alpha$ to be $\pi$. This implies that
the period of $\psi''$ is $6\pi$ and thus
the five-dimensional space is the orbifold
$S^5/\mathbb{Z}_2$. In other words, the mildly singular $D=11$
spaces that we constructed in section 5.1 of~\cite{GMSW} with
$a=c=1$ are in fact related, after dimensional reduction and
T-duality, to $S^5/\mathbb{Z}_2$.

\section{Discussion}

We have presented an infinite number of new Sasaki--Einstein
metrics of co-homo\-geneity one on $S^2\times S^3$, both in the
quasi-regular and irregular classes. As far as we know these are
the first examples of irregular metrics. As type IIB backgrounds
both classes should provide supergravity duals of a family of
${\cal N}=1$ superconformal field theories. Let us make some
general comments about the field theories. First we recall that
the geometries generically admit an $SU(2)\times U(1)^2$ isometry.
We also get additional baryonic $U(1)_B$ flavour symmetry factors
from reducing the RR four-form gauge potential on independent
three-cycles in $X_5$. Since here $X_5=S^2\times S^3$ this gives a
single $U(1)_B$ factor. Thus the continuous global symmetry group
of the dual field theory for all our examples is, modulo discrete
identifications,
\begin{equation}
   SU(2)\times U(1)^2\times U(1)_B~.
\end{equation}
From the geometry we can also identify the $R$-symmetry. It is generated
by the Sasaki Killing vector\footnote{This follows from the fact that the supercharges of the CFT are identified with the Killing spinors on $X_5$,
and these indeed are charged with respect to the canonical Sasaki
direction \cite{Gibbons:2002th}.} $\de/\de\psi'$ which is a linear
combination~\eqref{sasvec} of $\per\de/\de\alpha$ and $\de/\de\psi$
which generate the $U(1)^2$ isometry group.

This geometrical picture matches the discussion by
Intriligator and Wecht~\cite{Intriligator:2003jj} of
$R$-charges in ${\cal N}=1$ superconformal field theories.
Consider such a field theory with the above global symmetries.
In \cite{Intriligator:2003jj}
it is argued that the $R$-symmetry is not expected to mix
with the non-Abelian and
baryonic factors $SU(2)$ and $U(1)_B$. Thus it should be some
combination of the $U(1)^2$ factors precisely as we see in the
supergravity solutions. There are then two distinct possibilities. For
the quasi-regular metrics the Sasaki-Einstein
Killing vector generates a compact $U(1)_R$ symmetry in
the field theory and the $R$-charges of the fields must be rational. For
the irregular metrics, the $R$-symmetry is non-compact and we are
allowed irrational $R$-charges. Note that from~\eqref{per-pq}
and~\eqref{sasvec} we see that we have a relation between
the Killing vectors involving only quadratic algebraic numbers (by
which we mean they are \emph{square-roots} of rational numbers). This
implies that the $R$-charges are also quadratic algebraic
numbers, in agreement with the analysis in \cite{Intriligator:2003jj}.

In general, the AdS/CFT correspondence implies that the ratio of
the central charges associated to our spaces $Y^{p,q}$ to the
central charge of $\mathcal{N}=4$ super-Yang--Mills theory with
gauge group $SU(N)$, is equal to the ratio of the volumes of $S^5$
and $Y^{p,q}$~\cite{gubser}. This is simply given by the
right-hand side of the formula~\p{volume} without the $\pi^3$. If
we denote the central charge of the superconformal field theory
associated to $X_5$ as $a(X_5)$ we find
\be
   a(S^5)<a(T^{1,1})<a(S^5/\bZ_2)
      <a(T^{1,1}/\bZ_2)<a(Y^{2,1})<a(Y^{p,q})
\ee
for $(p,q)\ne (2,1)$. Since the volumes of our new spaces can be
arbitrarily small, the corresponding central charge can be
arbitrarily large. Assuming an $a$-theorem\footnote{For recent
progress in this direction, see \cite{kutasov}.}, we see that none
of the conformal field theories associated with $Y^{p,q}$ can
arise in the IR via RG flow of a perturbation of the known field
theories\footnote{It is interesting to
  observe that for any Einstein manifold $X_5\ne S^5$ with the same
Ricci  curvature as $S^5$, then $\vol(X_5)<\vol(S^5)$ (see section G of
  chapter 12 of \cite{besse}). Hence the field theories dual to a type
  IIB solution of the direct product form $AdS_5\times X_5$ should not
  arise as perturbations of ${\cal N}=4$ SYM theory.}
associated with $S^5$, $T^{1,1}$, $S^5/\bZ_2$ or $T^{1,1}/\bZ_2$.
This is to be contrasted with the field theory associated with
$T^{1,1}$ which has been argued to arise via a perturbation of the
field theory associated with the orbifold
$S^5/\bZ_2$~\cite{Klebanov:1998hh}.

In accordance with the $R$-charge discussion above it is simple to see,
from eq. (\ref{volume}),
that the central charges of the quasi-regular metrics that we have
constructed are rational, while the central charges of the irregular
metrics are irrational. In fact, the latter are quadratic algebraic
--  hence we find perfect agreement with the field theory prediction
of~\cite{Intriligator:2003jj}. We emphasize that irregular
Sasaki--Einstein manifolds are the only possible candidate duals
of a superconformal field theory with irrational $R$-charges and
hence central charges. In particular this requires that the
volumes of regular and quasi-regular Sasaki--Einstein metrics,
which give rational $R$-charges and hence central charges, must
always be rational fractions of that of $S^5$, with the same
curvature, which is always the case.
Indeed, if $X_5$ is a regular Sasaki-Einstein manifold, with
K\"ahler-Einstein base $B_4$ and with the $U(1)$ fibration being
that corresponding to the canonical bundle of $B_4$, we have (see
also \cite{bergman})
\bea
\mathrm{vol}(X_5)=\frac{\gamma(B_4)}{27}\mathrm{vol}(S^5)\label{ratio}\eea
where
\bea \gamma(B_4)=\int_{B_4}c_1^2\eea
is a Chern number of $B_4$, and $c_1\in H^2(B_4;\mathbb{Z})$ is
the first Chern class. In particular, note that $\gamma(B_4)$ is
always an integer. As an example, take
$B_4=\mathbb{C}P^1\times\mathbb{C}P^1$. Then $\gamma(B_4)=8$, and
(\ref{ratio}) gives the ratio $8/27$, which is indeed correct for
$T^{1,1}/\mathbb{Z}_2$ \cite{gubser}. The quasi-regular case is
more subtle, but a similar result still holds since now the Ricci
form (divided by $2\pi$) has rational, rather than integral,
periods: $c_1\in H^2(B_4;\mathbb{Q})$. Also note that similar
formulae hold in higher dimensions.

As mentioned in the introduction, these type IIB solutions
$AdS_5\times Y^{p,q}$ are T-dual to type IIA solutions which can
in turn be lifted to solutions of $D=11$ supergravity. Indeed, it
was the reverse route that originally led us to
$Y^{p,q}$~\cite{GMSW}. For the type IIA and $D=11$ solutions to be
good string or M-theory backgrounds it is necessary that they have
quantised fluxes. However, under T-duality this is simply a
consequence of the regularity of the spaces $Y^{p,q}$ (see for
example~\cite{Bouwknegt:2003wp}).

For the solutions to provide good {\it supergravity} duals to
superconformal field theories, $X_5$ should not have any small
cycles. As usual this requires that the volume of $Y^{p,q}$ is
chosen to be large, by an overall scaling the solution, and
corresponds to large 't~Hooft coupling $g_sN$ where $g_s$ is the
string coupling and $N$ is the number of D3-branes. However we
must also consider the size of the $\alpha$ circle.
From~\eqref{per-pq} it is clear that this can potentially be
small. However, we can always choose the overall scaling of the
solution so that there is a good type IIB supergravity
description. For example, consider the large $\lambda$ limit $p\gg
q$. We see that the overall volume of the manifold scales as
$1/p$. The period $\per$ of $\alpha$ scales as $1/q$ while $w(y)$
scales as $q^2/p^2$ and, hence, the length of the $\alpha$-circle
also scales as $1/p$. The dependence of the overall volume on $p$
is due solely to the size of the $\alpha$-circle; the other
dimensions are all of order one. Hence choosing $(g_sN)^{1/4}\gg
p$ all cycles are large and we have a good IIB supergravity
solution.  However, if we choose $p\gg(g_sN)^{1/4}\gg 1$ then the
$\alpha$-circle remains small and a good supergravity description
of the field theory will be provided by the T-dual type IIA
solution (or the $D=11$ solution when the IIA string coupling is
large).

It is interesting to note that under T-duality the canonical Sasaki
Killing direction associated with the $R$-symmetry does not get mapped
to the canonical Killing direction which generates the $R$-symmetry of
the $D=11$ solution. As shown in \cite{GMSW}, the latter is generated
by $\partial/\partial \psi$ and this always generates a compact
symmetry for the solutions we are discussing.
The discrepancy is proportional to the period $\per$ of
$\alpha$ which indicates that there are string
theory corrections to the $R$-symmetry when the $\alpha$-circle is small.

Beyond these general properties one would of course like to identify
the dual conformal field theories in more detail. They arise from
$N$ D3-branes at the tip of a singular Calabi--Yau geometry given by the
metric cone over $X_5$ with the field content determined by the
form of the singularity. It is suggestive to note that we can view our
$Y^{p,q}$ manifolds as a Lens space $L(p,1)$ fibred over a base $S^2$. The
Lens space is parametrised by the coordinates $(y,\psi,\alpha)$ in
the metric~\eqref{tinky} and the $S^2$ base by
$(\theta,\phi)$. Topologically $L(p,1)$ is the base of
the four-dimensional cone describing an $A_{p-1}$ singularity
$\bC^2/\bZ_p$. Thus perhaps one way to view the singular Calabi--Yau
geometry is as an $A_{p-1}$ singularity fibred over a collapsing $S^2$. The
Chern number $p$ describes the order of the singularity $A_{p-1}$ while
$q$ encodes the fibration over $S^2$.

If this is correct, one would then expect to find a related smooth
Calabi--Yau manifold by first resolving the base $S^2$ (in analogy
with the conifold) and then blowing up the singular $A_{p-1}$ fibres. This
resolution, and possibly others, should have important consequences
for the dual field theories and the generalisations obtained by adding
fractional branes. It would be of interest to construct these smooth
Calabi-Yau manifolds, which are expected to be co-homogeneity two.

\section*{Acknowledgements}

We thank Charles Boyer, Alex Buchel, Michael Douglas, Fay Dowker,
Joel Fine, Kris Galicki, Gary Gibbons, Jaume Gomis, Pierre
Henry--Labordere, Chris Herzog, Ken Intriligator, Rob Myers, Leopoldo
Pando Zayas, Joe Polchinski, Simon Salamon, Nemani Suryanarayana,
Richard Thomas, Alessandro Tomasiello and Arkady Tseytlin for helpful
discussions.
DM is funded by an EC Marie--Curie Individual Fellowship under contract
number HPMF-CT-2002-01539.
JFS is funded by an EPSRC mathematics fellowship. DW is supported
by the Royal Society through a University Research Fellowship.

\appendix

\section{The topology of the manifolds}
\label{top}

In this appendix we show that the total space $Y^{p,q}$ of the
$U(1)$ bundle over $S^2 \times S^2$ with relatively prime winding
numbers $p$ and $q$ over the two two-cycles is topologically
$S^2\times S^3$. This is a simple consequence of Smale's Theorem
\cite{smale} and the Gysin sequence for the $U(1)$ fibration. We
include the argument here for completeness.

Let $E$ be the complex line bundle over $S^2\times S^2$ with
winding numbers $p$ and $q$, where $(p,q)=1$. The boundary
$\partial E$ of $E$ is then our space $Y^{p,q}$. We denote the
projections $\Pi:E\rightarrow B$, $\pi:\partial E\rightarrow B$,
where $B=S^2 \times S^2$.

We first show that $Y^{p,q}$ is simply-connected. Consider the
following part of the long exact cohomology sequence for the pair
$(E,\partial E)$:
\bea \ldots \rightarrow H^4(E,\partial E; \mathbb{Z})
\stackrel{f}{\rightarrow} H^4(E;\mathbb{Z})
\stackrel{i^*}{\rightarrow} H^4(\partial E;\mathbb{Z}) \rightarrow
H^5(E,\partial E;\mathbb{Z})\rightarrow\ldots\eea
where $i:\partial E \rightarrow E$ denotes embedding and the map
$f$ forgets that a class is a relative class (has compact
support). The Thom isomorphism theorem states that
$H^*(B;\mz)\stackrel{\cup\Phi}\cong H^{*+2}(E,\partial
E;\mathbb{Z})$ for a complex line bundle $E$ over base $B$, where
the isomorphism is the cup product with the Thom class $\Phi\in
H^2(E,\partial E;\mz)\cong\mz$. Thus the last term of the above
sequence is
$H^3(B;\mz)=0$, since $B=S^2 \times S^2$.
Exactness of the sequence now implies
\bea H^4(\partial E;\mz)\cong H^4(B;\mz)/\left[c_1\cup
H^2(B;\mz)\right]\eea
%
where we have used the fact that
$H^2(B;\mz)\stackrel{\cup\Phi}{\cong} H^4(E,\partial E;\mz)$
and also that $f(\Phi)=\Pi^*(c_1)\in H^2(E;\mz)$.
Now $B$ is a deformation retract of $E$, so $H^4(E;\mz)\cong
H^4(B;\mz)\cong \mz$, and $H^2(B;\mz)\cong \mz\oplus\mz$ where the
generators are dual to the two $S^2$ cycles. Thus $c_1\cup
H^2(B;\mz)\subset \mz$. This subgroup is generated by all elements
of the form $pb + qa$ where $a,b\in\mz$, since $c_1$ is $p$ times
the first generator of $H^2(B;\mz)\cong \mz\oplus\mz$, and $q$
times the second. The subgroup is therefore $(p,q)\mathbb{Z}$,
where $(p,q)$ denotes highest common factor. Since the latter is
1, we find that $H_1(\partial E;\mathbb{Z})\cong H^4(\partial
E;\mz)=0$ is trivial. It follows that $\pi_1(\partial E)$ is also
trivial, so $\partial E$ is simply-connected.

One can similarly show that $H_2(\partial E;\mz)\cong H^3(\partial
E;\mz)\cong \mz$. For this we need the following part of the long
exact sequence:
\bea 0{\rightarrow} H^3(\partial
E;\mz)\stackrel{\delta^*}{\rightarrow} H^4(E,\partial
E;\mz)\stackrel{f}{\rightarrow}H^4(E;\mz)\rightarrow 0\eea
where we have now used the fact that $\partial E$ is
simply-connected. The homomorphism $\delta^*$ is the so-called
connecting homomorphism of the long exact sequence. The last two
terms are isomorphic to $\mz\oplus\mz$ and $\mz$, respectively.
First, note that $H^3(\partial E;\mz)$ can contain no torsion.
For, suppose that $nx =0$ for some $x\in H^3(\partial E;\mz)$ and
$n\in\mz$. Then $n\delta^*(x)=0$ as $\delta^*$ is a homomorphism.
But this implies that $\delta^*(x)=0$ as it is an element of
$\mz\oplus\mz$. But now exactness of the sequence implies that $x$
must be trivial. Finally, the rank of $H^3(\partial E;\mz)$ is
$\dim \mathrm{Im}\delta^* = \dim \mathrm{ker}f=1$.

Note also that $w_2(\partial E)=\pi^*w_2(B)=0$, where $w_2$
denotes the second Stiefel-Whitney class. Thus $\partial E$ is a
spin manifold.

We now use Smale's theorem \cite{smale}. This states that any
simply-connected compact 5-manifold which is spin and has no
torsion in the second homology group is diffeomorphic to $S^5\#
l(S^2\times S^3)$, for some non-negative integer $l$. Here $\#$
denotes connected sum, which means one excises small balls from
each manifold, and then glues the remaining sphere boundaries
together, with appropriate orientation. Clearly, taking the
connected sum of any manifold with $S^5$ gives back the same
manifold -- notice that $S^5\# l(S^2\times S^3)$, with $l=0$, is
simply $S^5$. It is straightforward to show\footnote{See exercise
15, Chapter 5 of \cite{maunder}.} that the second homology group
of $S^5\# l(S^2\times S^3)$ is $\mz^l$. Thus, from the analysis
above, we see that the topology of $\partial E=Y^{p,q}$ is
$S^2\times S^3$, for all $p,q\in \mz$ with $(p,q)=1$.


\end{document}